\documentclass[acmtog]{acmart}
\AtBeginDocument{%
  \providecommand\BibTeX{{%
    \normalfont B\kern-0.5em{\scshape i\kern-0.25em b}\kern-0.8em\TeX}}}

\usepackage{multirow}

\setcopyright{acmcopyright}
\copyrightyear{2022}
\acmYear{2022}

\acmConference[SIGGRAPH Asia 2022]{SIGGRAPH Asia 2022}{Dec 06--09,
  2022}{EXCO, Daegu, South Korea}
\acmPrice{15.00}
\acmISBN{978-1-4503-XXXX-X/18/06}

\acmSubmissionID{421}

\setcopyright{acmcopyright}\acmJournal{TOG}
\acmYear{2022}\acmVolume{41}\acmNumber{6}\acmArticle{200}\acmMonth{12}
\acmDOI{10.1145/3550454.3555501}

\begin{document}

\title{Reconstructing Personalized Semantic Facial NeRF Models From Monocular Video}

\author{Xuan Gao}
\authornote{This work was done when Xuan Gao, Chenglai Zhong and Jun Xiang were intern at Image Derivative Inc.}
\email{gx2017@mail.ustc.edu.cn}
\affiliation{%
	\institution{University of Science and Technology of China}
	\country{China}
}

\author{Chenglai Zhong}
\email{zcl2017@mail.ustc.edu.cn}
\affiliation{%
	\institution{University of Science and Technology of China}
	\country{China}
}

\author{Jun Xiang}
\email{xiangjunxjkd1@mail.ustc.edu.cn}
\affiliation{%
	\institution{University of Science and Technology of China}
	\country{China}
}

\author{Yang Hong}
\email{hymath@mail.ustc.edu.cn}
\affiliation{%
	\institution{University of Science and Technology of China}
	\country{China}
}

\author{Yudong Guo}
\email{guoyudong@idr.ai}
\affiliation{%
	\institution{Image Derivative Inc}
	\country{China}
}

\author{Juyong Zhang}
\email{juyong@ustc.edu.cn}
\authornote{Corresponding author (\href{mailto:juyong@ustc.edu.cn}{juyong@ustc.edu.cn}).}
\affiliation{%
	\institution{University of Science and Technology of China}
	\country{China}
}

\authorsaddresses{Authors' addresses:
	$\{$Xuan Gao, Chenglai Zhong, Jun Xiang, Yang Hong, Juyong Zhang$\}$, University of Science and Technology of China, 96 Jinzhai Road, Hefei 230026, Anhui, China, $\{$gx2017@mail.ustc.edu.cn, zcl2017@mail.ustc.edu.cn, xiangjunxjkd1@mail.ustc.edu.cn, hymath@mail.ustc.edu.cn, juyong@ustc.edu.cn$\}$; Yudong Guo, Image Derivative Inc, 998 Wenyi West Road, Hangzhou, Zhejiang, China, guoyudong@idr.ai.}

\renewcommand{\shortauthors}{Gao, et al.}

\begin{abstract}

We present a novel semantic model for human head defined with neural radiance field. The 3D-consistent head model consist of a set of disentangled and interpretable bases, and can be driven by low-dimensional expression coefficients. Thanks to the powerful representation ability of neural radiance field, the constructed model can represent complex facial attributes including hair, wearings, which can not be represented by traditional mesh blendshape. To construct the personalized semantic facial model, we propose to define the bases as several multi-level voxel fields. With a short monocular RGB video as input, our method can construct the subject's semantic facial NeRF model with only ten to twenty minutes, and can render a photo-realistic human head image in tens of miliseconds with a given expression coefficient and view direction. With this novel representation, we apply it to many tasks like facial retargeting and expression editing. Experimental results demonstrate its strong representation ability and training/inference speed. Demo videos and released code are provided in our project page: https://ustc3dv.github.io/NeRFBlendShape/
\end{abstract}



\begin{teaserfigure}
\includegraphics[width=\textwidth]{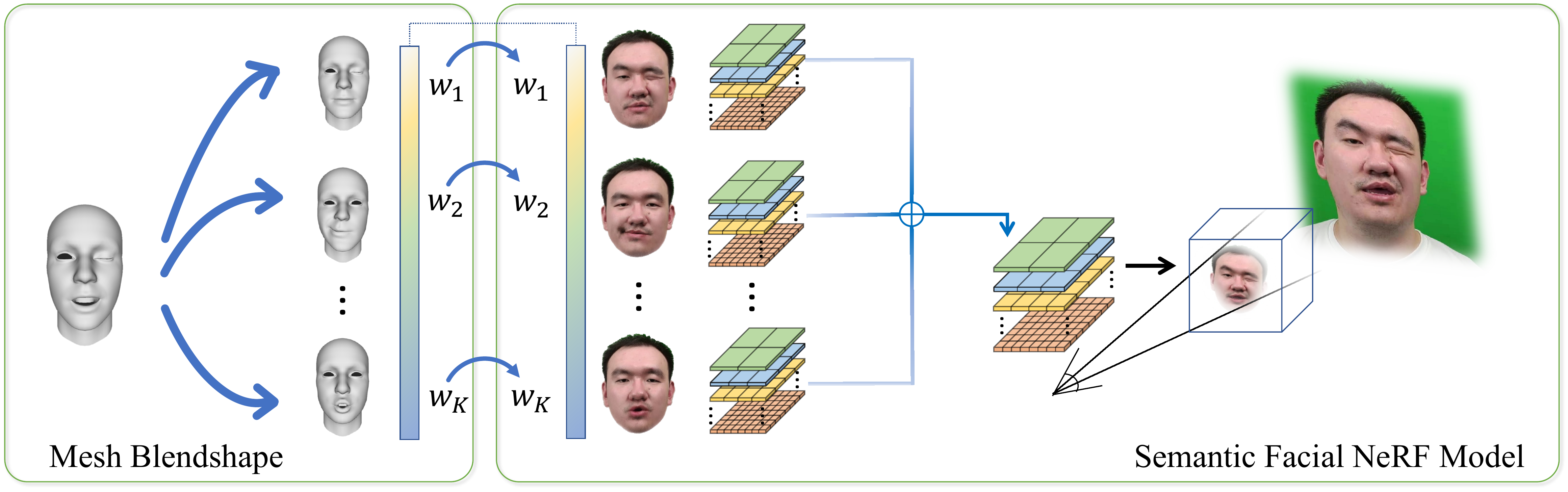}
\caption{A semantic model for human head defined with neural radiance field is presented. In this model, multi-level voxel field is adopted as basis with corresponding expression coefficients, which enables strong representation ability on the aspect of rendering and fast training.}
\Description{figure description}
\end{teaserfigure}
\maketitle

\section{Introduction}
3D face/head representation is an important research problem in computer vision and computer graphics, and has wide applications in AR/VR, digital games and movie industry. How to represent the dynamic head and 
faithfully reconstruct a personalized human head model from a monocular RGB video is an important and challenging research topic. With the hypothesis that human head could be embedded into a low dimensional space. Parametric semantic head models, like blendshape, have been studied and improved for a long time. The blendshape head model, in the form of linear/bilinear combination of different facial expressions, has the following advantages. It is a semantic parameterization. The combination coefficients have intuitive meaning for the users as the strength or influence of specific facial expressions. Meanwhile, the blendshape constructs a reasonable shape space which can help the user freely control and edit in the space.

The generalized semantic head models like FaceWarehouse~\cite{cao2013facewarehouse} aim to model different subjects with different expressions, and thus may ignore personalized geometry and texture details. To construct a personalized blendshape model, traditional mesh based methods usually adopt deformation transfer\cite{sumner2004deformation,li2010example,garrido2016reconstruction} and multilinear tensor-based 3DMM\cite{vlasic2006face,cao2014displaced,cao2018stabilized}. However these methods usually have the following disadvantages. First, mesh based parametric models are hard to represent personalized non-face parts like hair and teeth. Second, to use an RGB supervision, we have to use approximated differentiable rendering techniques to alleviate the non-differentiable problems. Third, deformation transfer cannot reconstruct expressions realistically due to limited representation ability. Last, facial expressions are characterized by many factors such as ages and muscle movements, and these factors are hard to be accurately expressed by predefined blendshapes.

Recently, NeRF based methods have made it possible to synthesize photorealistic images. Some works integrate NeRF with GANs\cite{Chan2021pi,gu2021stylenerf,niemeyer2021giraffe,schwarz2020graf,zhou2021cips,chan2022efficient,deng2021gram}. However, this kind of generative models couple expression, identity and appearance together, resulting that the expressions can not be easily controlled. HeadNeRF\cite{hong2021headnerf} proposes to disentangle different semantic attributes, but it could not represent personalized facial dynamics and facial details due to its generic model capacity. AD-NeRF\cite{guo2021adnerf} and NerFACE\cite{Gafni_2021_CVPR} could generate highly personalized facial animation, their user-specific training make the model learn more personalized facial details. However, they need a long time to train a reasonable dynamic head field. According to our experiments in section \ref{sec::concate}, this is because they concatenate the expression condition with Fourier positional information and directly input it to the MLP. Both the Fourier positional encoding and the ``concatenate'' strategy is not ideal for fast training. The Fourier encoding is not friendly for MLP for fast convergence. And the concatenation operation does not contain any combination law to discover the relation between local and global features (in NeRF case, positional information and expression condition). Therefore, it takes a long time for MLP to learn how to use the expression condition to predict RGB and density.

Recently, local features have been explored to improve NeRF's quality and efficiency. The original NeRF's local feature is the Fourier positional encoding, which takes a long time to converge. Following works designs different kinds of local features to improve NeRF. Some methods adopt a voxel field to accelerate the training process\cite{sun2021direct,yu_and_fridovichkeil2021plenoxels}. Other works use the voxel field to accelerate ray marching and volume rendering\cite{yu2021plenoctrees,garbin2021fastnerf,liu2020neural}. EG3D\cite{chan2022efficient} adopts a compact and efficient tri-plane architecture enabling geometry-aware synthesis. TensoRF\cite{Chen2022ECCV} factorizes the 4D scene tensor into multiple compact low-rank tensor components to separate local features. Among these methods, instant neural graphics primitives (INGP)\cite{mueller2022instant} demonstrated a remarkable performance improvement in both training and rendering. It uses a highly compressed compact data structure, multi-level hash table, to make it possible to store a multi-level voxel field. A novel design of INGP is that the feature query collision is solved in an adaptive way. Features in different levels could be trained together. Together with the help of high-performance ray-marching implementation, it could train a static NeRF scene less than 1 min and render one frame in tens of milliseconds.

Although a lot of methods have been proposed to speed up the training and inference of a static NeRF field, it still remains a problem to achieve a fast training of a dynamic scene such as the complicated head deformation. As our baseline shows, using a direct ``concatenate'' strategy to combine local features and expression code together as the input of MLP, which is very common for NeRF based head application, is not efficient and sufficient to model dynamic head motions.

In this paper, we present a personalized semantic facial model architecture defined on multi-level voxel field. It not only inherits the semantic meaning of mesh blendshape used for tracking, but also has more personalized facial detailed attributes especially for no-face part. Each basis of our model is a radiance field of a specific expression, represented by a multi-level voxel field. We adopt the multi-resolution hash tables to store the multi-level features for performance consideration. For any novel expressions, it can be expressed as the weighted combination of voxel bases with the expression coefficients. We adopt an MLP to interpret the voxel field as a radiance field for volume rendering. To further accelerate the ray marching in volume rendering and make the optimization focus on the region possibly occupied by head, we design an expression-aware density grid update strategy. Thanks to powerful representation ability and fast convergence of our implicit model, our method outperforms other similar head construction methods in both modeling quality and construction speed. Our method can construct a photo-realistic personalized semantic facial model in around 10-20 minutes, which is remarkably faster than related NeRF based head technique. As our model is trained from a video of a specific person and combines the features in a latent space, it could capture personalized details including non-linear deformation (cheek folds, forehead wrinkle) and user-specific attributes(mole, beard). Compared with traditional mesh based blendshape models, our model can be constructed from a short RGB video and generate high-fidelity view consistent head images with different expressions.

In summary, the contributions include the following aspects:
\begin{itemize}
\item We present a novel semantic model for human head defined with neural radiance field. Our constructed NeRF basis not only has a disentangled semantic meaning, but also embodys more personalized facial attributes including muscle actions and detailed texture. Therefore, the constructed digital avatars can model facial motions quite well and generate photo-realistic results.

\item Our representation combines multi-level voxel fields with expression coefficients in the latent space. The multi-resolution features could efficiently learn head details in different scales. The linear blending design could modulate the local features in advance to adapt to MLP's input distribution, which makes our model cost much less time to construct and express more realistic facial details.

\item With this novel representation, digital human head related applications like facial reenactment can be easily achieved and have remarkable performance, which implies its potential usage in photo-realistic animation industry.




\end{itemize}

\section{Related Work}

\paragraph{Parametric Head Model}
Under the hypothesis that human head shape space can be well disentangled as identity, expression and appearance, Blanz and Vetter proposed 3DMM\cite{blanz1999morphable} to embed 3D head shape into several low-dimensional PCA spaces. Mesh based parametric head model has been further studied by a lot of following works. To improve its representation ability, some work extends it to multilinear models\cite{cao2013facewarehouse,vlasic2006face}, non-linear models\cite{ranjan2018generating,tran2018nonlinear,guo20213d} and articulated models with corrective blendshape to improve its modeling ability\cite{li2017learning}. Both mesh based methods and deep learning based methods have been widely used in many related applications. However, mesh based parametric models usually can not represent personalized facial details due to its limited representation ability. Meanwhile, existing mesh based parametric models can not represent non-face parts especially for hair. Some works handle this problem using deformation transfer\cite{cao2016real,garrido2016reconstruction,hu2017avatar,ichim2015dynamic,sumner2004deformation} or neural network\cite{bai2021riggable,chaudhuri2020personalized,yang2020facescape} to get user-specific blendshape basis. 

To break through the limited representation ability of explicit mesh based digital human representation, many works adopt the implicit representation to improve the model capacity and visual quality~\cite{hong2021headnerf,zhuang2021mofanerf,yenamandra2021i3dmm,zheng2022avatar,Gafni_2021_CVPR,jiang2022selfrecon,wang2021prior}. i3DMM\cite{yenamandra2021i3dmm} is the first neural implicit function based 3D morphable model of full heads.  HeadNeRF\cite{hong2021headnerf} proposes a generic head parametric model based on neural radiance field. Although neural implicit function based representations have demonstrated strong representation ability, a generic model often still lacks personalized facial details.
NerFACE\cite{Gafni_2021_CVPR} presents a personalized NeRF based human head model. However, their method requires a long time for training and inference for each subject.
IM Avatar\cite{zheng2022avatar} presents an implicit LBS model. Note that both our method and IM Avatar have an implicit blendshape architecture. The main difference is that IM Avatar focuses on detailed geometry and appearance and our model focuses more on photorealistic rendering and efficient training/inference. Another difference is that IM Avatar uses a backward non-rigid ray marching to find the canonical surface point for each ray. Our ray marching is performed in the deformed space.

\paragraph{Human Portrait Synthesis}
Many methods have been proposed for facial reenactment and novel view synthesis. Image based methods\cite{Siarohin_2019_NeurIPS,zakharov2020fast,Pumarola_ijcv2019} adopt warping fields or encoder-decoder architectures to synthesize the images. As these methods represent the 3D deformation in the 2D space, artifacts may appear for large pose and expression changes. Morphable model\cite{kim2018deepvideo,thies2019deferred,thies2020neural,thies2016face,JuyongZhang2020FacialER} based methods use a parametric 3D model to synthesize a digital portrait. Deep Video Portraits\cite{kim2018deepvideo} uses rendered correspondence maps together with an image-to-image translation network to output photo-realistic imagery. Deferred Neural Rendering\cite{thies2019deferred,thies2020neural} proposes an object-specific neural textures which can be interpreted by a neural renderer.

\paragraph{Neural Radiance field} NeRF\cite{mildenhall2020nerf} proposes to represent a scene with an MLP and utilize the volume rendering for novel view synthesis task. As NeRF is differentiable, its inputs can be only multi-view images. Due to the above listed characteristics, NeRF has been widely used to 3D geometry reconstruction\cite{wang2021neus,yariv2021volume}, 4D scene synthesis \cite{park2021nerfies,li2021neural,park2021hypernerf} and digital human modeling\cite{peng2021neural,peng2021animatable,weng2022humannerf},etc. Besides, a lot of research focus on improving NeRF's representation ability\cite{barron2021mip} and reducing the number of inputs\cite{chibane2021stereo,niemeyer2022regnerf,yu2021pixelnerf}.

Recently, NeRF has also demonstrated its strong representation ability in human head modeling. Many works adopt NeRF to represent dynamic human head scene, and synthesis high-fidelity 3D consistent result. Generative head models~\cite{Chan2021pi,gu2021stylenerf,niemeyer2021giraffe,schwarz2020graf,zhou2021cips,chan2022efficient,deng2021gram} use latent code to generate the rendering result. Although they usually have a good pose control over the result, but do not support expression editing due to its generative adversarial training strategy. Generic parametric head model \cite{hong2021headnerf,zhuang2021mofanerf} disentangles latent space of human head as identity, expression and appearance space, and to some extent realize semantical control over head transformation. However, generic head model often ignores personalized facial details and user-specific facial muscle movements due to limited MLP capacity. AD-NeRF\cite{guo2021adnerf} and NerFACE\cite{Gafni_2021_CVPR} are subject-specific models, and it can generate high fidelity human head animation controled by voices or expressions. However, both AD-NeRF and NerFACE need days for training and seconds for inference. And we found both of them tend to learn a smooth head scene and sometimes ignore high-frequency facial attributes.

\paragraph{Voxel Representation for NeRF Acceleration}
With the help of voxel field, NeRF could spare its training burden across the local features, which will significantly improve the training speed\cite{sun2021direct,yu_and_fridovichkeil2021plenoxels}. Voxel field could also help store the spacial information like density distribution in advance to accelerate the inference speed\cite{yu2021plenoctrees,garbin2021fastnerf,liu2020neural,lombardi2021mixture}. Recently, instant neural graphics primitives\cite{mueller2022instant} adopts multi-level hash table to augment a shallow MLP and achieves a combined speedup of several orders of magnitude. It can train a static scene with NeRF using only several seconds, and render the scene in tens of milliseconds. However, these methods could not be directly used for dynamic scenes due to its complex non-rigid deformations. Meanwhile, it is hard to perform ``pruning'' operation for voxel grid in dynamic scenes, which is usually important for ray-marching and information storage.

\section{Method}

\begin{figure*}[t]
  \centering
  \includegraphics[width=0.97\linewidth]{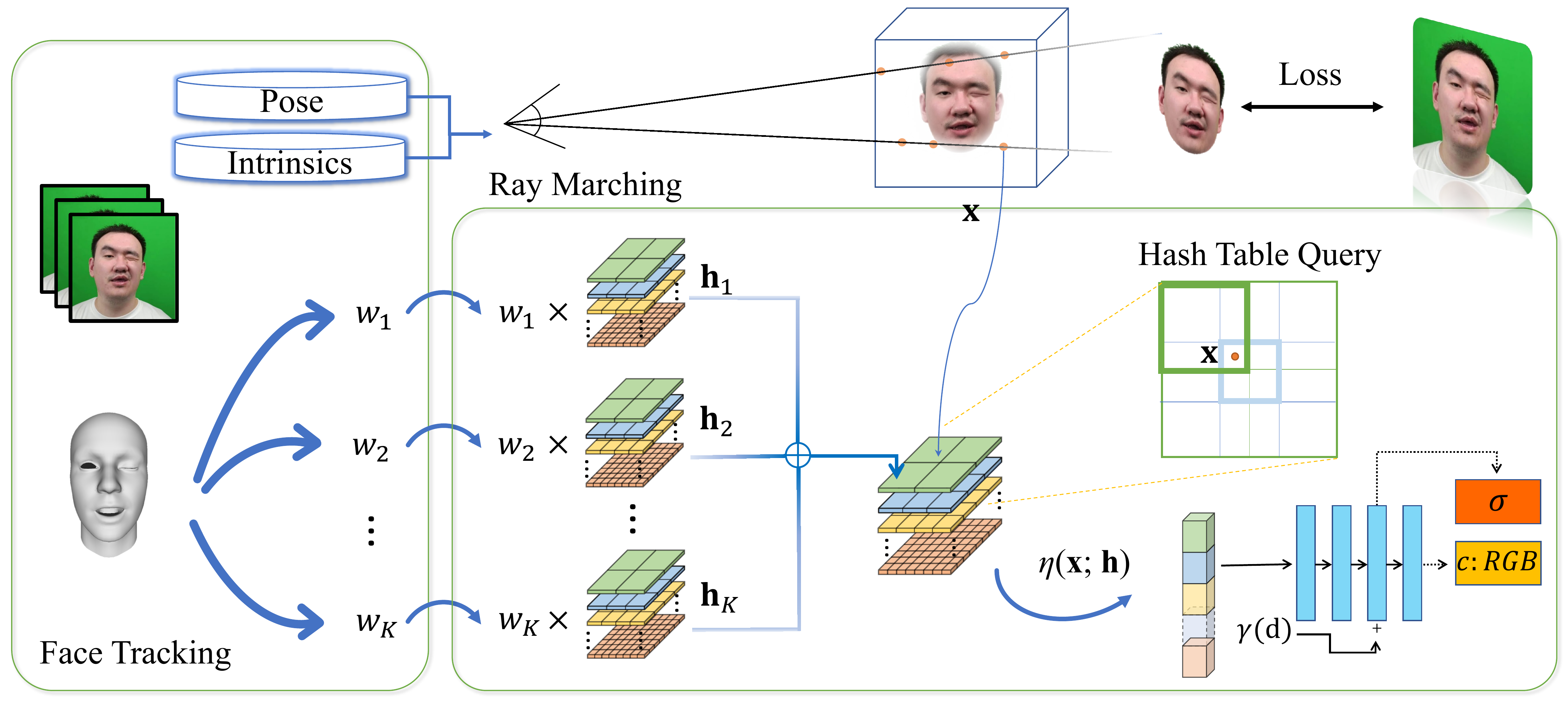}
  \caption{Our pipeline, we track the RGB sequence and get expression coefficients, poses and intrinsics. Then we use the tracked expression coefficients to combine multiple multi-level hash tables to get a  hash table corresponding to a specific expression.  Then the sampled point is queried in hash table to get voxel features, we use an MLP to interpret the voxel features  as RGB and density. We fix the expression coefficients and optimize the hash tables and MLP to get our head model.  }
  \label{fig:pipeline}
\end{figure*}

In this work, we propose a novel personalized human head representation that takes a series of specially designed neural radiance fields as the bases of the human head. Similar with traditional mesh blendshape like FaceWarehouse~\cite{cao2013facewarehouse} , each basis of the proposed model has a specified semantic meaning, such as eye closed and jaw forward, which makes it easy for users to use a low-dimensional code to generate desired human head images. 

Our head model linearly combines multi-level voxel fields  with expression coefficients in the latent space, and the multi-resolution features could efficiently learn head details in different scales. Our
linear blending design could modulate the local features in advance to adapt to MLP’s input distribution, which reduces the training burden of MLP and expresses
more realistic facial details. We improve the rendering efficiency of our model by concentrating the sampling near surfaces. With these designs, the proposed method could construct a set of NeRF bases in less than 20 minutes. Meanwhile, the trained model has interactive rendering speed and can render photo-realistic human head images. Furthermore, the results generated by our personalized NeRF-based blendshape can be further semantically edited, such as freely adjusting camera parameters and independently changing the subject's expression to any desired expression while keeping other attributes unchanged. An overview of the proposed representation is shown in Fig.~\ref{fig:pipeline}, and the algorithm details will be presented in the following.

\subsection{NeRF based Linear Blending Representation}


Similar with\cite{Gafni_2021_CVPR,hong2021headnerf,chan2022efficient}, our representation is also based on neural radiance field~\cite{mildenhall2020nerf}, and we represent it by the MLP-based implicit function. Besides, we associate expression bases with multi-level hash tables~\cite{mueller2022instant} and endow each hash table with specified semantic attributes via an elaborately designed training strategy. We denote the representation of our model as $\mathcal{R}$ and formulate it as:
\begin{center}
  \begin{equation}
    \begin{aligned}
    I = \mathcal{R}_{\theta}(C, \mathbf{h_0} + \mathbf{H} {\mathbf{w}}),
    \label{equ:blendshape_representation}
    \end{aligned}
  \end{equation}
\end{center}
where $\theta$ indicates the learnable weight parameters of the MLP. $C$ is the camera parameter used for rendering, including the extrinsic
and intrinsic matrices. $\mathbf{h_0} \in \mathbb{R}^{L \times T \times F}$ is the multi-level hash table representing the mean shape of the blendshape in latent space, where $L$ is the number of hash table's levels, $T$ is the hash table size, and $F$ is the number of feature dimensions per entry of the hash table. $\mathbf{H} = \{\mathbf{h}_1, \mathbf{h}_2, \ldots, \mathbf{h}_{K}\}, \mathbf{h}_i\in \mathbb{R}^{L \times T \times F}$ is the multi-level hash table representing the expression displacement basis in latent space. $K$ is the number of the expression bases of our model, and $\mathbf{w} = \{w_1, w_2, \ldots, w_K\} \in \mathbb{R}^{K}$ is the expression coefficient. $I$ is the human head image rendered by $\mathcal{R}_{\theta}$ according to the above parameters.

\subsection{Rendering}
The rendering process of our model is shown in the right part of Fig.~\ref{fig:pipeline}. We first calculate the corresponding hash tables for a given expression coefficient $\mathbf{w}$ as:
\begin{equation}
  \mathbf{h} = \mathbf{h_0} + \mathbf{H} \mathbf{w} = \mathbf{h_0} + \sum_{i=1}^K{w_i \mathbf{h}_i}, \mathbf{h} \in \mathbb{R}^{L \times T \times F},
  \label{equ:blendshape_hashtable}
\end{equation} 
where we combine the bases in multi-level voxel space instead of combining the blendshape basis in explicit space as mesh blendshape. Then we adopt the model architecture of \cite{mueller2022instant} and formulate the MLP-based implicit function $g_{\theta}$ of NeRF as:
\begin{equation}
  g_{\theta}: (\eta(\mathbf{x}; \mathbf{h}), \gamma(\mathbf{d})) \mapsto (\sigma, c),
  \label{equ:implicit_func}
\end{equation} 
where $\mathbf{x} \in \mathbb{R}^{3}$ is a 3D point sampled from one ray. $\mathbf{d} \in \mathbb{R}^{3}$ indicates a unit vector representing view direction. $\eta(\mathbf{x}; \mathbf{h}) \in \mathbb{R}^{L \cdot F}$ is the encoding of $\mathbf{x}$ about $\mathbf{h}$, and it is obtained by linearly interpolating the feature vectors indexed by the hash value of $\mathbf{x}$'s transformed integer corner points~(please refer to~\cite{mueller2022instant} for details). $\gamma(\mathbf{d})$ is the positional encoding of $\mathbf{d}$, which projects $\mathbf{d}$ onto the first 16 coefficients of the spherical harmonics basis. $\sigma$ and $c$ denote the predicted density and color of $\mathbf{x}$, respectively.

Actually, the expression coefficients could combine the local features encoded by the multi-level hash tables, and this effectively enhances the representation ability of our model. Thus, we can use a lightweight MLP to represent the implicit function $g_{\theta}$. The network architecture of $g_{\theta}$ is a 4  layers MLP with 64 neurons width. It means that the rendering of our model can be executed efficiently. Lastly, we generate the rendered image $I$ by the following volume rendering:
\begin{center}
  \begin{equation}
    \begin{aligned}
    &I(r)=\int_{0}^{\infty}p(t) c(r(t)) dt, 
    \\
    \text{where}~\quad &p(t)=exp(-\int_{0}^{t}\sigma(r(s))ds)\sigma(r(t)).
    \label{equ:volume_rendering}
    \end{aligned}
  \end{equation}
\end{center}
$r(t)$ represents a ray emitted from the camera origion. The head mask can be generated in a similar way:
\begin{equation}
    M(r)=\int_{0}^{\infty}p(t) dt.
    \label{equ:mask_render}
\end{equation}

In summary, the rendering process of our model consists of the following steps: First, we use the expression coefficients to linearly combine the multi-level hash tables that represent different expression bases in latent space. Then we cast rays to get sampled points. By querying these points in the combined hash table, the hash encoding of the sample point w.r.t the combined hash table is generated. We further use a lightweight MLP-based implicit function to map the generated hash encoding to $RGB$ and density for volume rendering. We want to point out that these steps can be fast calculated, and we will further skip empty space when building our model. Therefore, the rendering of our model is efficient, and the constructed personalized semantic head model can quickly generate the target head image with the given target expression coefficient. Besides, efficient rendering actually speeds up the construction of our models as well.


\subsection{Construction}
The proposed personalized NeRF-based semantic facial model can be constructed using only a 3-5 minutes monocular RGB video thanks to the above-mentioned elaborate design. In the following, we will present the algorithm details to construct our model.
\subsubsection{Data Preprocessing.} 
\label{sec:data_proc}
Firstly, we use an existing mesh-based facial blendshape~\cite{cao2013facewarehouse} to track the face in the input video similar to \cite{guo2018cnn}. Then we can obtain the expression coefficients and the head pose parameters of each frame. Like HeadNeRF~\cite{hong2021headnerf}, we take the human head pose parameters as the extrinsic camera parameter of the corresponding frame, which implicitly aligns the underlying geometry of each
frame to the same spatial location. Lastly, we randomly extract $N$ frames from the input video to train our model, and the head mask of the selected frame is generated by the off-the-shelf segmentation methods\cite{yu2018bisenet} . We denote the optimized expression coefficients of selected frames as $\{{\mathbf{w}}^1, {\mathbf{w}}^2,\ldots, {\mathbf{w}}^N\}$.
\subsubsection{Training.} The learnable variables of our model include the network parameters $\theta$ of the implicit function $g_{\theta}$ and the multi-level hash tables ($\mathbf{h_0}, \mathbf{H}$) representing expression bases. Meanwhile, we take $\{\mathbf{w}^1, \mathbf{w}^2, \ldots, \mathbf{w}^N\}$ as the expression coefficients of our personalized NeRF-based blendshape and freeze them while training our model. 
The loss terms used to train our model include the following three terms:

\paragraph{Photometric Loss.} This loss requires that the rendered result is consistent with the input RGB image, which is common for NeRF-based reconstruction and can be formulated as:
\begin{equation}
  L_{color} = \sum_{r \in \mathcal{S}} \|I(r) - I_{GT}(r)\|_{1},
\end{equation}
where $\mathcal{S}$ is the set of rays in each batch, and $I(r)$, $I_{GT}(r)$ are the predicted RGB colors and ground truth for ray $r$ respectively.

\paragraph{Mask Loss.} This loss requires that the rendered mask of Eq.~\eqref{equ:mask_render} is consistent with the ground truth head mask. It is formulated as:
\begin{equation}
  L_{mask} = \sum_{r \in \mathcal{S}} \|M(r) - M_{GT}(r)\|_{1},
\end{equation}
where $M(r)$ and $M_{GT}(r)$ are the predicted mask value and ground truth for ray $r$ respectively. This loss makes sure density outside the head region is zero, and also lets the head region become an opaque object quickly.

\paragraph{Perceptual Loss.} The perceptual loss $L_{LPIPS}$ of \cite{zhang2018unreasonable} is utilized to provide robustness to slight misalignments and shading variations and improve details in the reconstruction.
We choose VGG as the backbone of LPIPS. As LPIPS uses convolutions to extract features, we sample $B$ patches with size $W \times W$, and render a total of $B \times W \times W$ rays in each batch~($B$ is also the number of frames in each batch). The rendered patch is compared against the patch with the same position on the input image. Similar strategy is used in \cite{weng2022humannerf,schwarz2020graf}.

In summary, the overall loss of training our model is defined as:
\begin{equation}
  L_{total} = \lambda_{1}L_{color}+\lambda_{2}L_{mask} + \lambda_{3}L_{LPIPS},
\end{equation}
where $\lambda_i$ is a scalar for balancing different terms. To minimize the total loss, we propose a well designed training strategy, and it contains three steps:

\begin{figure}[t]
  \centering
  \includegraphics[width=0.9\linewidth]{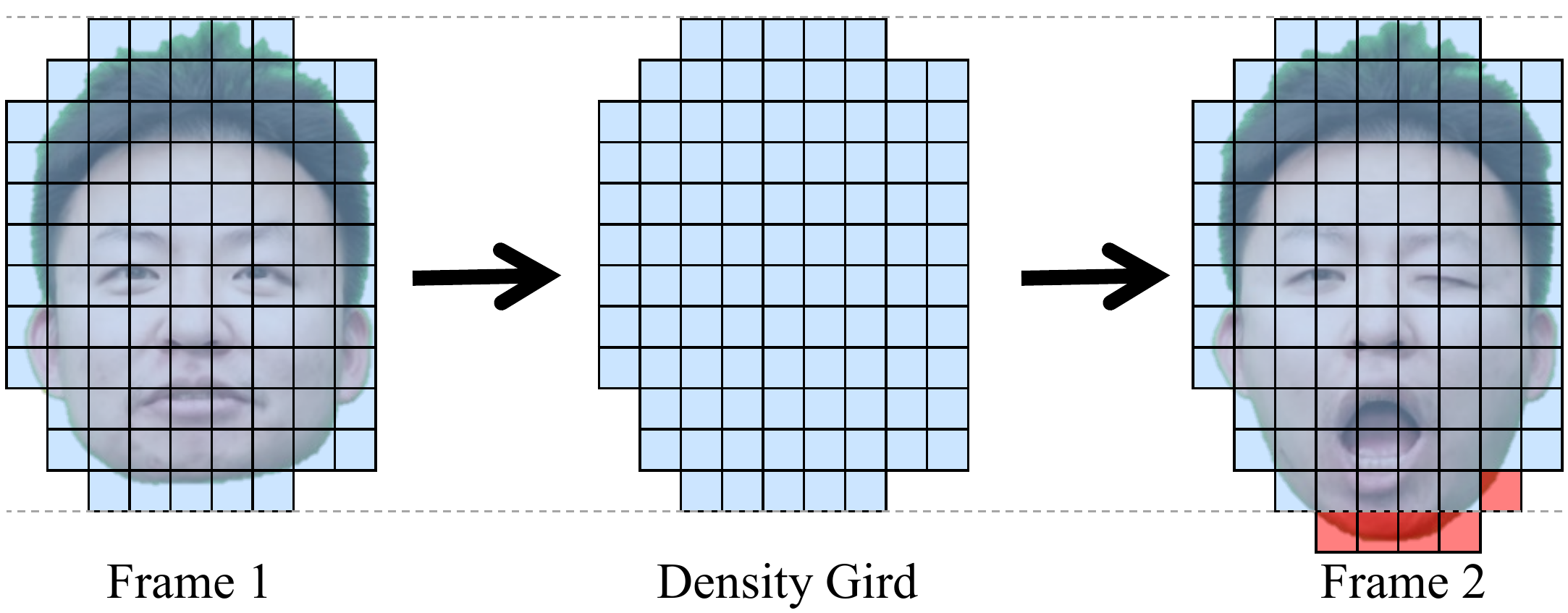}
  \caption{The density grid of a specific expression may not cover all the expression cases. Heads in some frames may be out of the range.}
  \label{fig:density_cover}
\end{figure}

\begin{figure*}[htb]
  \centering
  \includegraphics[width=0.9\linewidth]{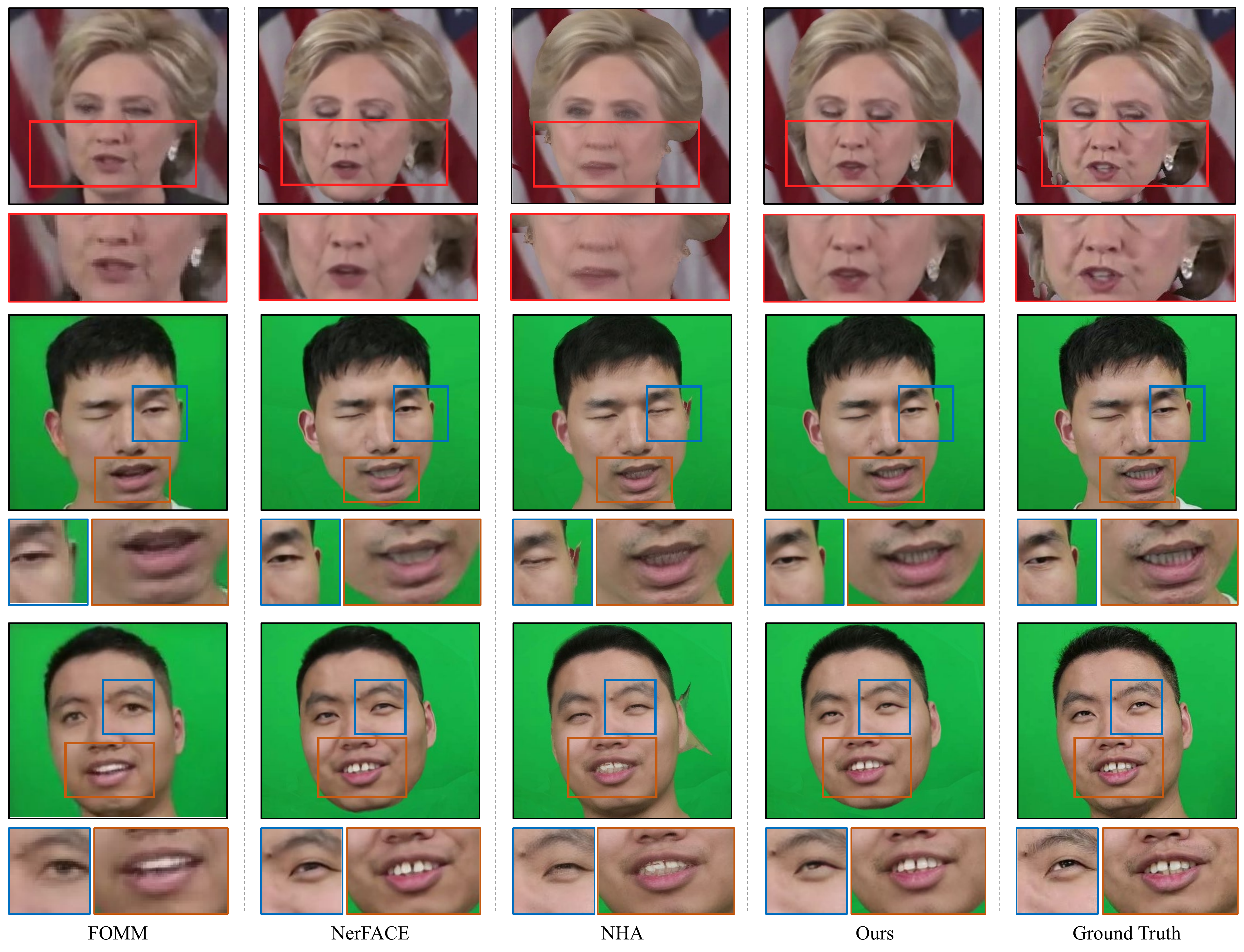}
  \caption{Comparison with state-of-the-art head modeling and facial reenactment methods. We can see that our model reconstruct high-fidelity expressions and facial details. YouTube ID of  Hillary Clinton's video is -yHgE9W699w.}
\label{fig:qualitative_evaluation}
\end{figure*}

In the first two epochs. we set  \begin{math}\lambda_{1}\end{math}, \begin{math} \lambda_{2}\end{math} to 1 and \begin{math}\lambda_{3}\end{math} to 0. The mask loss could make the model quickly learn the distribution of 3D density because it is a supervision directly on the density distribution. Stable density distribution is also good for the following density grid update.

Although the mask loss could help the convergency of density field, we found the parsing mask is not very accurate especially for the hair part. Therefore, for 2nd-7th epoch, we set \begin{math}\lambda_{2}\end{math}, \begin{math}\lambda_{3}\end{math} to zero and use photometric loss only. We use RGB data to supervise the NeRF training to learn fine detailed color and geometry.


Lastly, we not only randomly sample rays, but also sample patches. For randomly sampled rays, we set \begin{math}\lambda_{1}\end{math} to 1 and \begin{math}\lambda_{2},\lambda_{3}\end{math} to 0, for sampled patches, we set $\lambda_{1}$ to 0.1 and  $\lambda_{3}$ to 0.1 . We sample patches in mouth part in 1/2 probability and sample patches across the whole image otherwise.

\subsubsection{Expression-Aware Density Grid Update}

We use a \begin{math}128^3\end{math} density grid to store local density information to instruct ray-marching to skip empty space. Different from the static scene considered in~\cite{mueller2022instant}, the captured dynamic human head is changing over time with different poses and expressions. As the example shown in Fig.~\ref{fig:density_cover}, the density field of a specific expression could not cover all expression cases. We should not use the density field of a certain expression to determine the density grid. In our implementation, we compute density grid of each basis as:

\begin{equation}
  \begin{aligned}
    \hat{\mathbf{h}}_i = \mathbf{h}_0 + \hat{w_i}\mathbf{h}_i,
    ~\quad~\quad~\quad
    \\
    \textrm{where} ~\quad \hat{w}_i = \max_{j \in [1,N]} {w}_i^j.
  \end{aligned}
\end{equation}
${w}_i^j$ is the $i$-th element of ${\mathbf{w}}^j$. We compute the element-wise maximum of the density grids of all $\hat{\mathbf{h}}_i$ to get the final density grid to approximate the natural 3D range of head expression.

\subsubsection{Discussion on Fast Training}
Our model can converge quickly due to the following reasons:

Firstly, the relationship between global condition and local features (in our case, expression coefficients and queried positional features) have been considered in our architecture. This linear blending architecture could reduce the burden of MLP in learning how to use expression information to transfrom the positional information. We also validate our linear blending is much better than a direct "concatenate" strategy. See \ref{sec::concate} for experimental details.

Secondly, our model could capture multi-scale details at the same time. Features in different levels of multi-resolution hash table could be jointly optimized.

Thirdly, ray marching steps in empty space are skipped. we use a density grid to conduct efficient ray sampling. The density grid update considers every expression action to make sure every voxel occupied by the head region are considered.

\section{Experiments}

\subsection{Implementation Details}
 We implement our semantic facial model with Pytorch~\cite{paszke2019pytorch}, and the CUDA extension of Pytorch is employed to implement the raymarching and volume rendering parts. Our model is trained with Adam solver~\cite{kingma2014adam} with batch size 4. The dimension of our expression coefficients is 46, and we set the patch size of the perceptual loss to 32. The parameters of our hash table are listed in Tab.~\ref{tab:hash_para}. All the results are tested on one RTX 3090 card.
 
Some training videos are from the datasets collected by Neural Voice Puppetry\cite{thies2020nvp} and SSP-NeRF\cite{liu2022semantic}. Since the expressions in these videos are mainly normal talking styles which are not very challenging, we didn't do evaluation on the whole dataset. We capture some monocular videos with exaggerated expressions and large head rotations. In each captured video, the subject is asked to perform arbitrary expressions, and we have got the permissions from these subjects for research purposes. For each target person, we collect a 3-5 min monocular RGB video with $512 \times 512$ resolution and 30 fps. Thus we captured 5000-10000 frame images for a single person. The last 500 frames serve as the testing dataset, and we randomly extract 3000-4000 frames from the rest as our training dataset. 


\begin{figure}[t]
  \centering
  \includegraphics[width=0.95\linewidth]{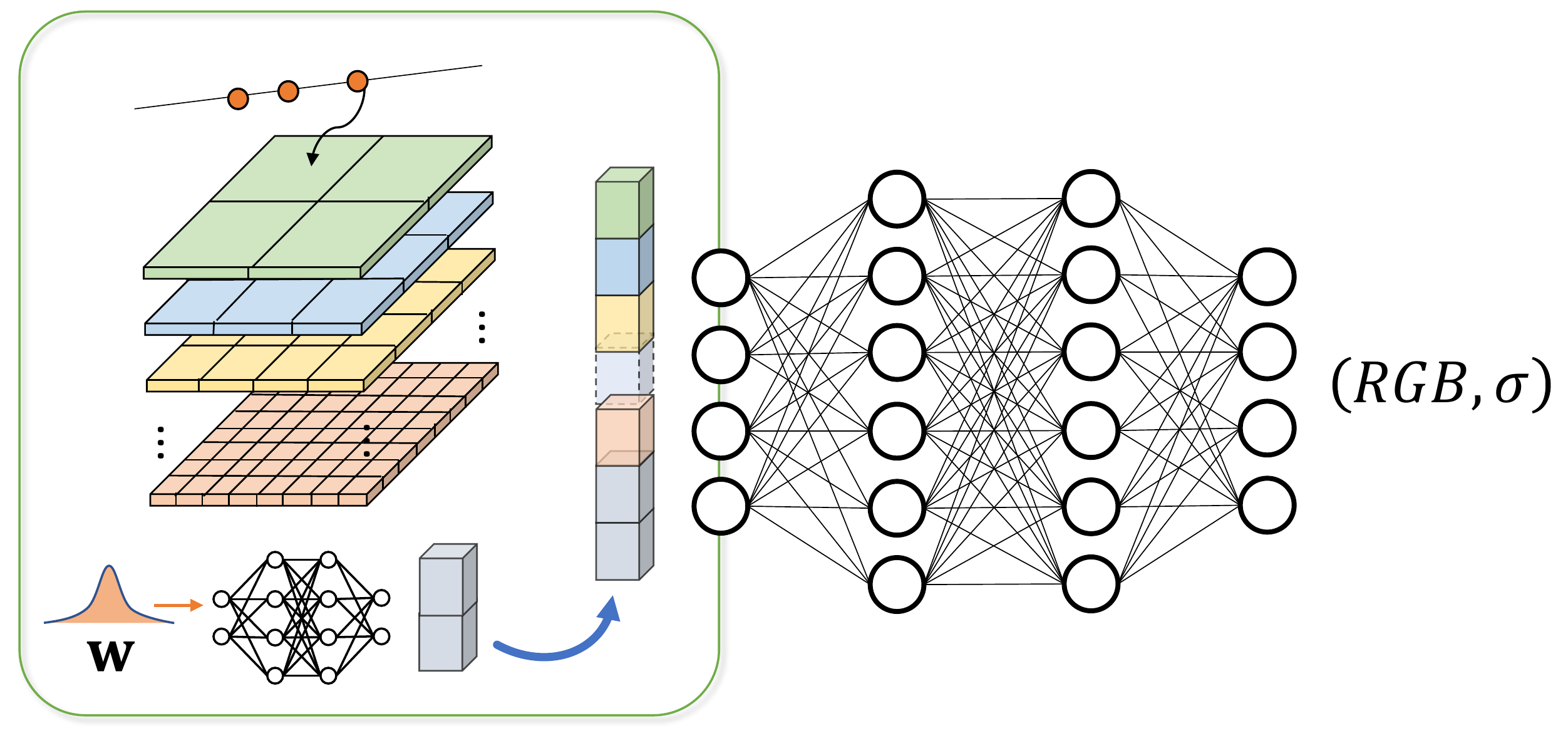}
  \caption{Baseline architecture. The queried feature in hash table is concatenated with expression code as the input of MLP. We use a deeper and wider MLP to demonstrate its representation ability. A 2-layer MLP is used to map the expression coefficients to be concatenated with the queried feature.}
  \label{fig:baseline}
\end{figure}

\begin{table}[t]
  \caption{The parameters of hash table}
  \label{tab:freq}
  \begin{tabular}{ccl}
    \toprule
    Parameter&Value\\
    \midrule
    Number of levels& 16\\
    Hash table size& $2^{14}$\\
    Number of feature dimensions per entry & 4\\
    Coarsest resolution& 16\\
    Finest resolution& 1024 \\
    Initial distribution& \begin{math} U(-10^{-4},10^{-4}) \end{math} \\
  \bottomrule
\end{tabular}
\label{tab:hash_para}
\end{table}


\subsection{Comparison}
In this part, we compare our model's rendering quality and representation ability with existing state-of-the-art facial reenactment or head modeling methods. Specifically, FOMM~\cite{Siarohin_2019_NeurIPS} uses a reference image and a driving video as inputs to generate a motion video sequence. NerFACE~\cite{Gafni_2021_CVPR} and Neural Head Avatars~(denoted as NHA)~\cite{grassal2022neural} use the same training data as ours. NerFACE is NeRF-based head modeling, which takes the concatenation of the expression code and positional encoding information as input. Neural Head Avatars explicitly reconstruct the full head on a FLAME model.

Fig.~\ref{fig:qualitative_evaluation} shows the qualitative comparison between our model and the above methods. It can be observed that our model is superior to others. We found the 3D consistency of FOMM is not as good as others. This drawback may originate from its 2D CNN architecture without enough 3D space knowledge. NerFACE tends to learn low-frequency signals and struggles to model head detailed information. Neural Head Avatars have more personalized details than NerFACE. However, the explicit mesh domain restricts its representation ability. As shown in this figure, undesired geometric artifacts appear in NHA's results. In contrast, our results are most consistent with the ground truth, and the fine-level details of the human head can be represented and rendered.

Tab.~\ref{table:quantitative_evaluation} shows the quantitative comparison between our model and other methods, where the Mean Squared Error~(MSE), L1 distance, Peak
Signal-to-Noise Ratio~(PSNR), Structure Similarity Index~(SSIM)~\cite{wang2004image} and Learned Perceptual Image
Patch Similarity~(LPIPS)~\cite{zhang2018perceptual} are computed. It is worth
noting that our method outperforms these existing methods in terms of any reconstruction error, which further verifies the effectiveness and superiority of our model.

\begin{figure*}[h]
  \centering
    \includegraphics[width=\linewidth]{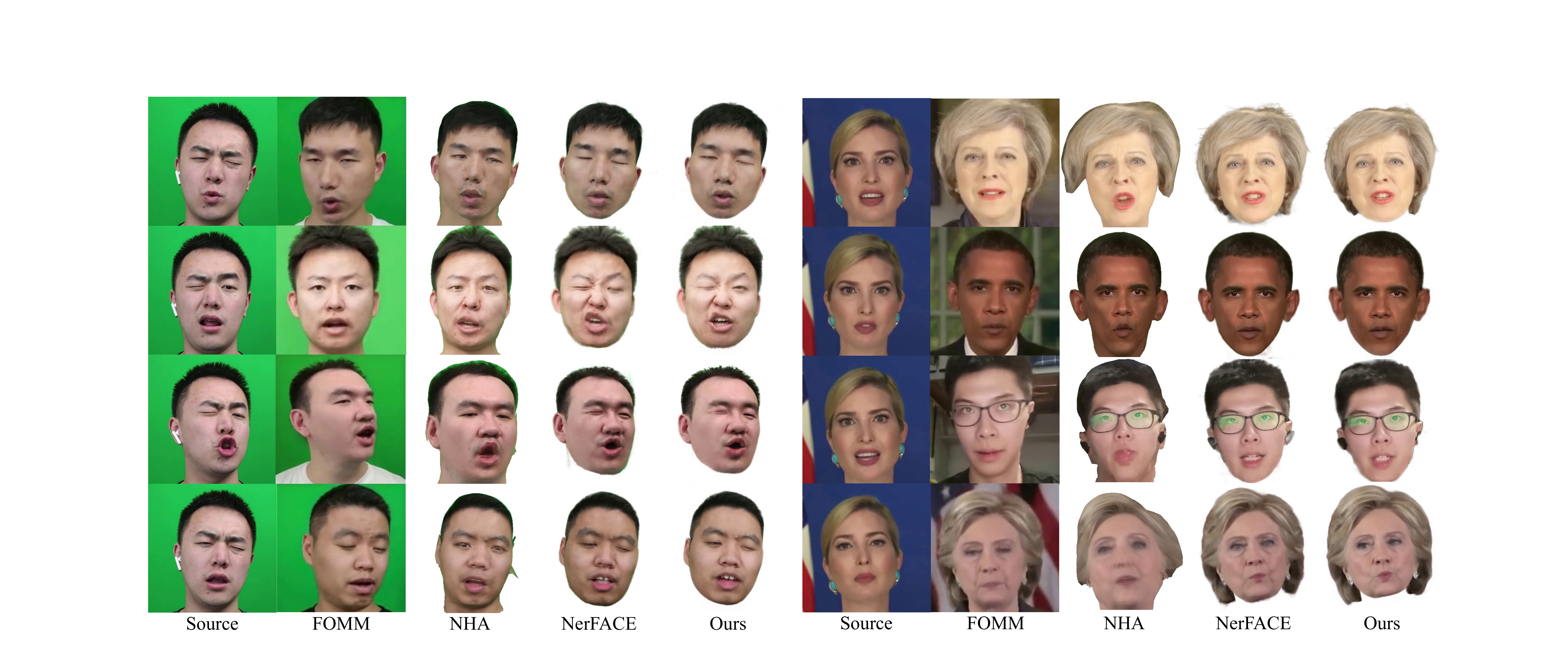}
  \caption{Our model can be easily used for facial reenactment. Compared with other methods, our method also demonstrates more
personalized facial details and synthesizes a more reasonable human head conditioned on expression coefficients in cross-identity
reenactment. YouTube ID of Ivanka Trump's video is -dNU9e0SYIg. YouTube ID of Theresa May's video is nOj49CzODEU. YouTube ID of Barack Obama's video is IQJW4\_FvVKo. YouTube ID of Hillary Clinton's video is -yHgE9W699w.}
  \label{fig:reenact}
\end{figure*}

As the expression coefficients space used for tracking is disentangled from identity space, we could use anyone's facial expression coefficients of the mesh-based blendshape to combine our bases and render the radiance field with a given view. Based on this observation, we apply our model to perform facial reenactment where we transfer the facial expressions from one person to another. Specifically, we track the source subjects' video and get poses, camera intrinsics, and expression coefficients. The poses and intrinsics are utilized to cast rays, and we further use expression coefficients to combine targets' bases. Then the sampled point is queried in the final hash table and interpreted by MLP to predict RGB and density. Compared with other methods, our method also demonstrates more personalized facial details and synthesizes a more reasonable human head conditioned on expression coefficients in cross-identity reenactment~(See Fig.~\ref{fig:reenact}), which demonstrates that our bases are semantically correct and suitable for reenactment.

\begin{figure*}[htb]
  \centering
  \includegraphics[width=0.95\linewidth]{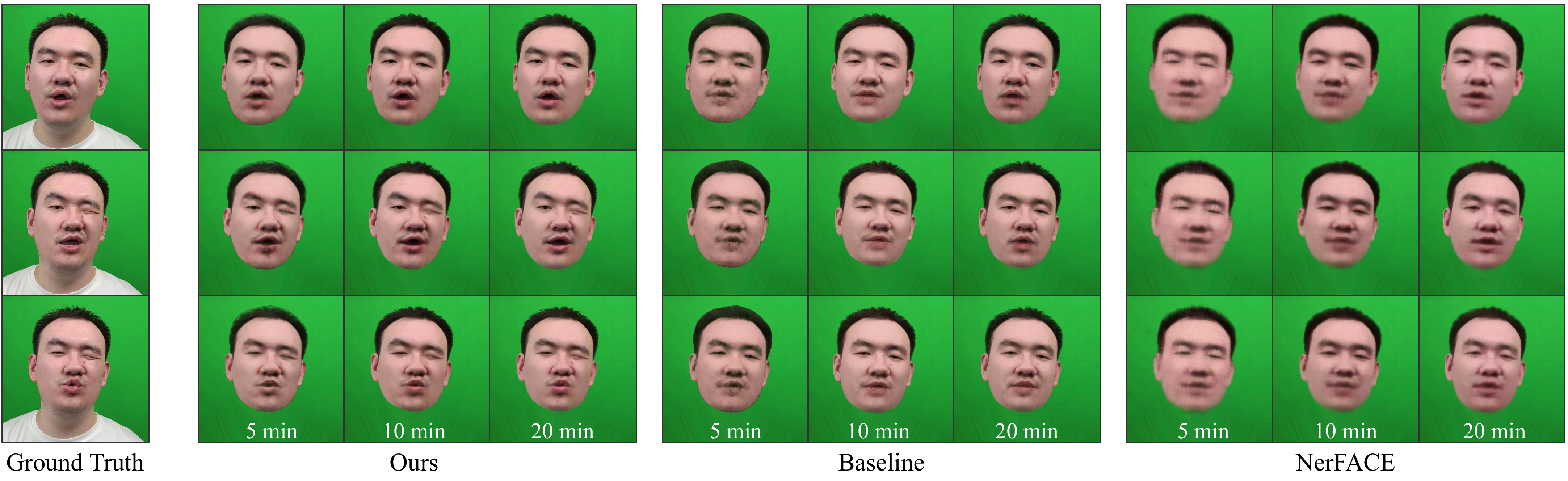}
  \caption{Comparison with directly ``concatenate'' strategies including our baseline implementation and NerFACE~\cite{Gafni_2021_CVPR}.} 
  \label{fig:image_training}
  \Description{baseline}
\end{figure*}

\begin{figure}[htb]
  \centering
  
  \includegraphics[width=\linewidth]{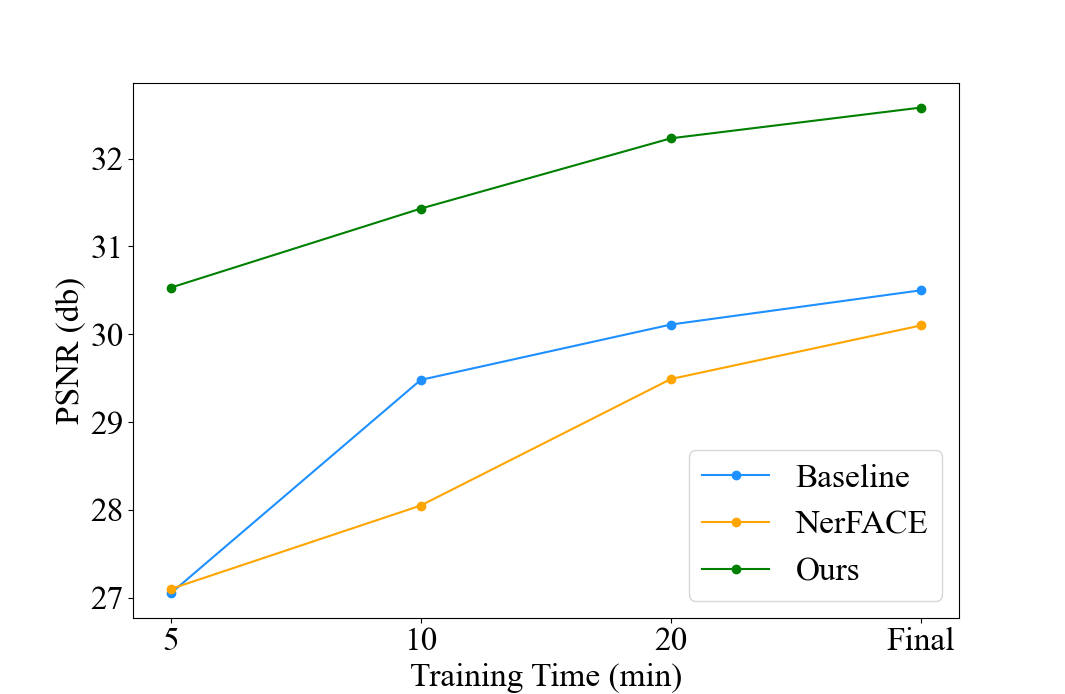}
  \caption{Comparison with baseline and NerFACE on training speed.}
  \label{fig:comparePSNR}
\end{figure}


\begin{table}[t]
\centering
\caption{Quantitative evaluation of our method in comparison to state-of-the-art facial reenactment methods based on self-reenactment. We compute the mean value and standard deviation of every method.}
\begin{tabular}{ccccc}
Metrics & FOMM & NerFACE & NHA & Ours \\
\hline
{MSE($10^{-3}$)$\downarrow$} 
 & 1.75(0.81) & 0.75(0.45) & 0.69(0.54) & \textbf{0.48}(0.32) \\
\hline
{L1($10^{-2}$)$\downarrow$} 
 & 1.87(0.52) & 0.84(0.30) & 0.80(0.29) & \textbf{0.70}(0.23) \\
\hline
{PSNR$\uparrow$} 
 & 28.32(2.45) & 32.24(2.70) & 32.85(3.01) & \textbf{34.15}(2.58)\\
\hline
{SSIM($10^{-1}$)$\uparrow$} 
 & 9.29(0.28) & 9.67(0.15) & 9.69(0.16) & \textbf{9.73}(0.13) \\
\hline
{LPIPS($10^{-2}$)$\downarrow$} 
 & 5.30(1.73) & 3.50(1.64) & 3.37(1.88) & \textbf{2.67}(1.32)\\
\hline
\end{tabular}
\label{table:quantitative_evaluation}
\end{table}

\begin{figure*}[htb]
  \centering

   \includegraphics[width=0.9\linewidth]{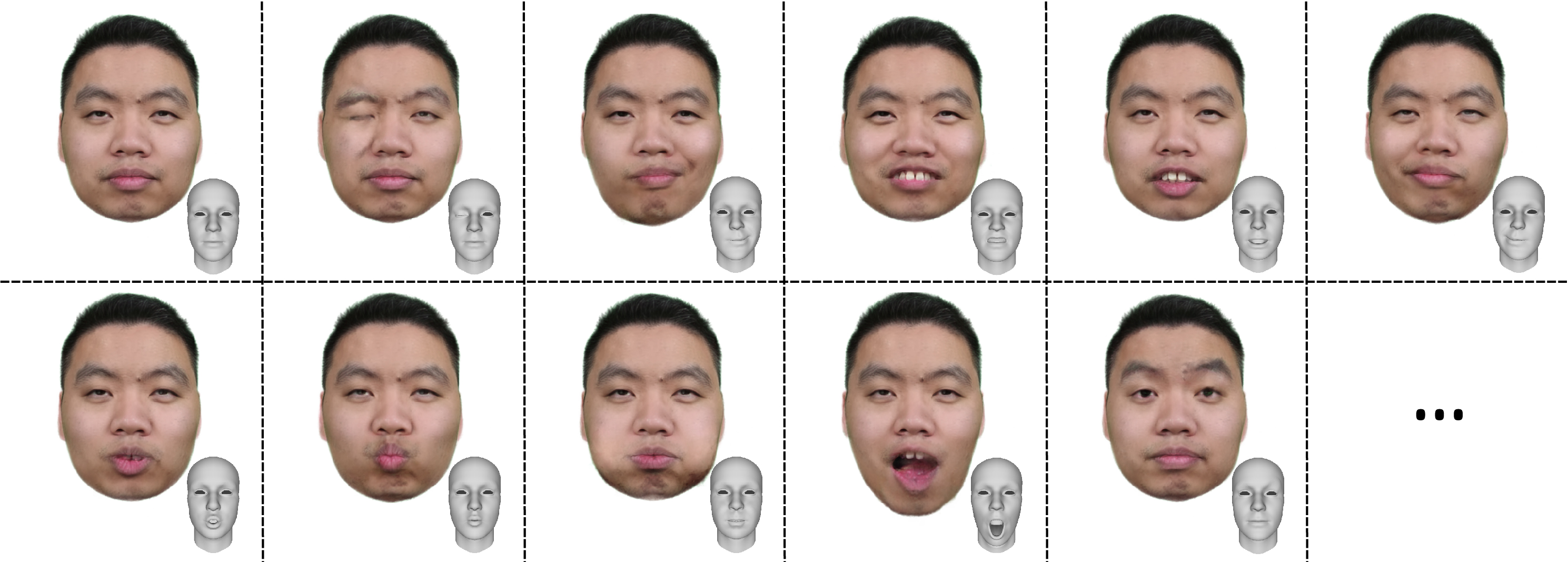}
  \caption{Our bases are consistent with mesh blendshapes on the aspect of semantical meaning but with a photo-realistic rendering quality. Our bases contain much more subject-specific facial details.}
  \label{fig:nerfbasis}
\end{figure*}

\subsection{Comparison with ``Concatenate'' Operation}
\label{sec::concate}
We want to point out that the multi-level voxel field is not the only reason for our model's representation ability and training efficiency. In fact, the implicit linear blending architecture also plays an important role in the learning process of our model. To verify this, we design a ``concatenate'' baseline model. This baseline use only one hash table, and the queried feature is concatenated with expression coefficients as the input of the MLP-based implicit function. Note that the ``concatenate'' operation is frequently used in many NeRF-based head models\cite{Gafni_2021_CVPR,guo2021adnerf,hong2021headnerf}. In addition, We keep the training strategy the same and use a much deeper MLP (7 layers to predict density and 1 layer to predict RGB, width 128) for this baseline to predict the RGB and density of sampled points. Fig.~\ref{fig:baseline} shows the architecture of this baseline. Besides, we note that NerFACE~\cite{Gafni_2021_CVPR} is also a ``concatenate'' model, where the concatenation of the expression coefficients and the Fourier positional encoding is taken as the input of the MLP-base implicit function. Thus we regard NerFACE as another baseline of our method.

As shown in Fig.~\ref{fig:image_training}, our model could learn a dynamic head scene in less than 20 minutes, in which time neither the baseline nor NerFACE could get any plausible result. Meanwhile, we find that both baseline and our method could quickly learn the rigid part of the human head. The difference is that the baseline tends to learn the rigid region first and then the dynamic region, such as the eyes and mouth. In contrast, our model could learn much more dynamic relationships in a limited short time. As shown in this figure, our model could faithfully reconstruct facial expressions like the open mouth and closed eyes after training for only 5 minutes. The comparison implies that a simple ``concatenated'' input is hard for MLP to learn the facial dynamic region no matter whether a voxel feature is used. The reason of our superiority may be that our implicit blendshape architecture enables local features to be modified by expression coefficients in a global manner, which makes the local features adapt to MLP's input distribution and reduces the learning burden of our model's MLP. Fig.~\ref{fig:comparePSNR} depicts the relationship between PSNR and training time. We can see that our model's training efficiency significantly outperforms both baseline and NerFACE.

\subsection{Geometry Visualization}

\begin{figure}[htb]
  \centering
  \includegraphics[width=0.95\linewidth]{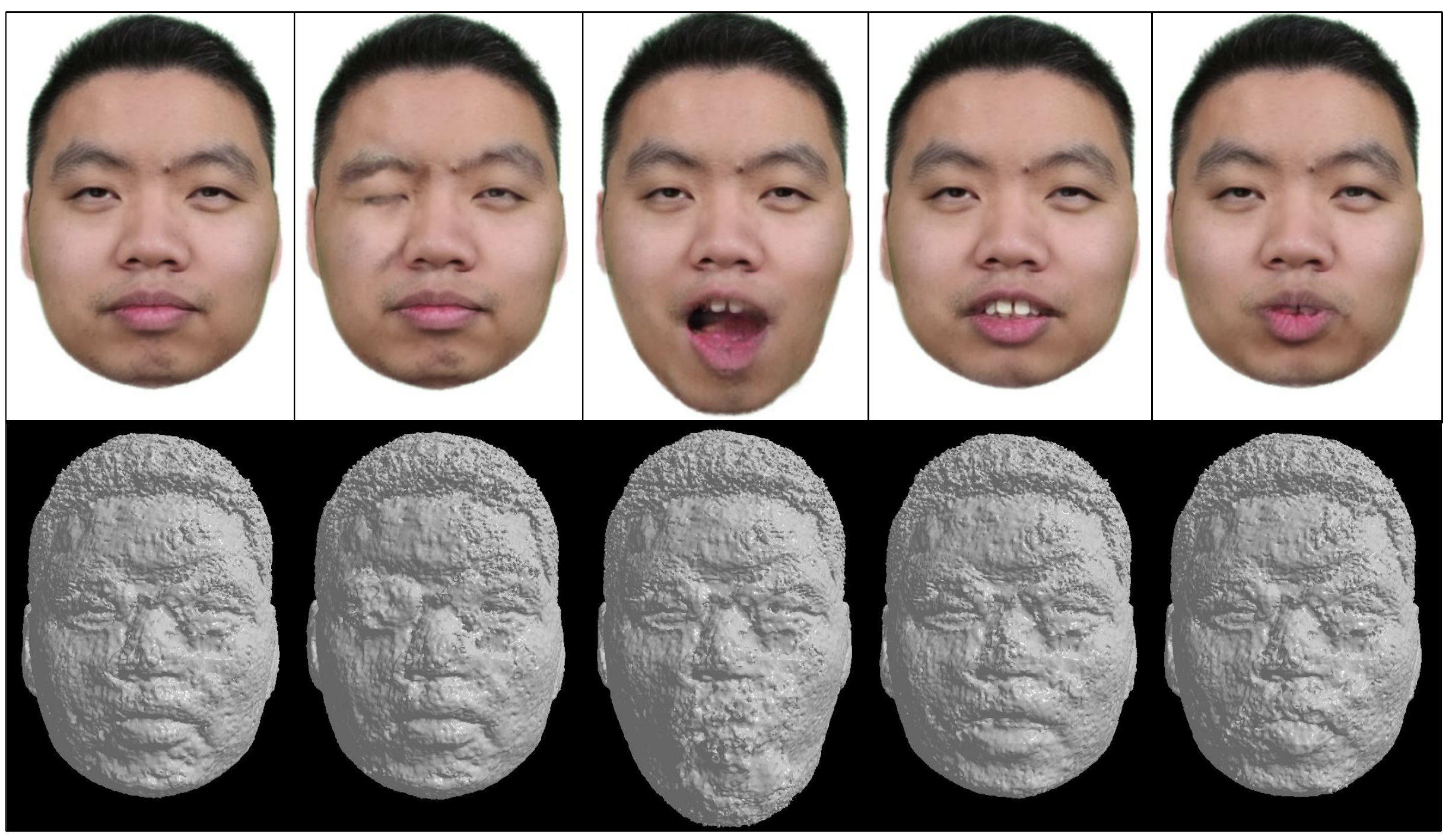}
  \caption{We extract the iso-surface from the density field with marching
cubes. Our model
could learn the geometric attributes like eyes, nose
and detailed hair thanks to the radiance field background.}
  \label{fig:geo}
\end{figure}

We extract the iso-surface from the density field with marching cubes. Although our model focuses more on photorealistic rendering and efficient training/inference, as shown in Fig.~\ref{fig:geo}, we found that the geometry is reasonable. Note that we don't use any direct supervision on the geometry like normals or depths, but our model could still learn the geometric attributes like eyes, nose and detailed hair thanks to the radiance field representation.
This also implies the 3D consistency of our head model which ensures robust novel view synthesis in Fig.~\ref{fig:3d_consist}. We also find noise occured on the extracted surface, and this may be circumvented by using state-of-the-art NeRF based surface representation like \cite{wang2021neus,yariv2021volume} and we leave it as a future work.

\subsection{Bases Visualization}
The semantical meaning of our hash table basis is inherited from the mesh blendshape used for tracking, see Sec.~\ref{sec:data_proc} . Fig.~\ref{fig:nerfbasis} demonstrates some selected expression bases of our model and the mesh-based blendshape, where semantical correspondence between our bases and the mesh-based blendshape are shown. We can find that the expressions of our bases are consistent with the mesh-based blendshape's expressions. Differently, the results generated by our model have higher rendering quality and highly personalized facial attributes. Moreover, our bases can represent various personalized facial details like hair and moles. The  subject-specific habits, such as the wrinkles and folds movements, could also be seen on each basis of our model. 

\subsection{Novel View Synthesis}
\begin{figure}[htb]
  \centering

  \includegraphics[width=\linewidth]{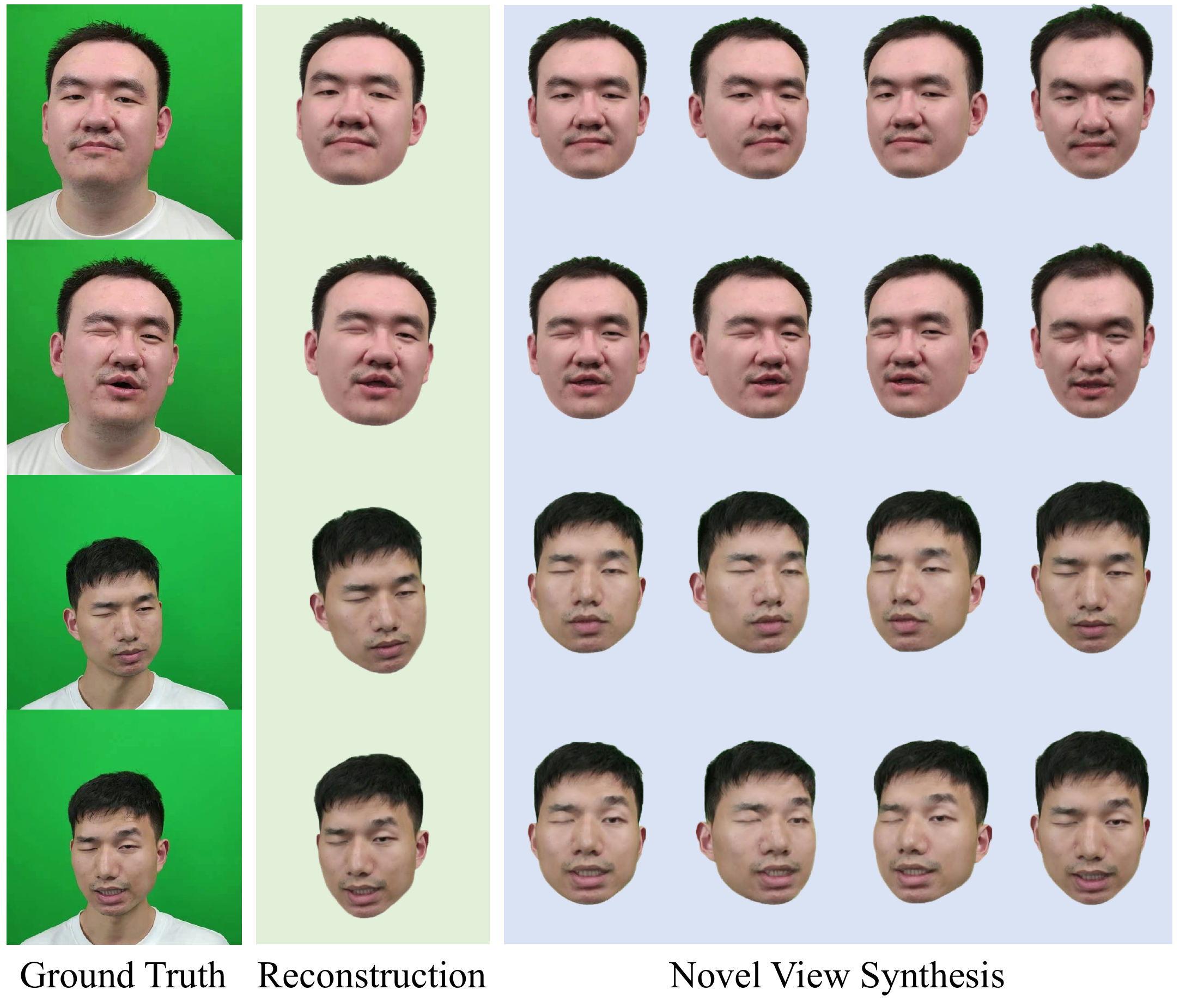}
  \caption{Our model is a 3D consistent representation, and thus the view direction can be freely adjusted}
\label{fig:3d_consist}
\end{figure}

Our model also disentangles the camera parameters (Equ.~\eqref{equ:blendshape_representation}). Thus we can freely adjust the camera parameters of our model to generate target results with any desired rendering view. Fig~\ref{fig:3d_consist} shows the novel view synthesis application based on our model. We first use a set of expression coefficients from the testset to combine bases to get the corresponding radiance field. Then the rendered images with different rendering views are generated by the volume rendering. Thanks to the 3D consistency of NeRF, these rendered results have remarkable multi-view consistency. 

\section{Ablation Study}



\subsection{Discussion on Perceptual Loss}
\begin{figure}[htb]
  \centering
  \includegraphics[width=\linewidth]{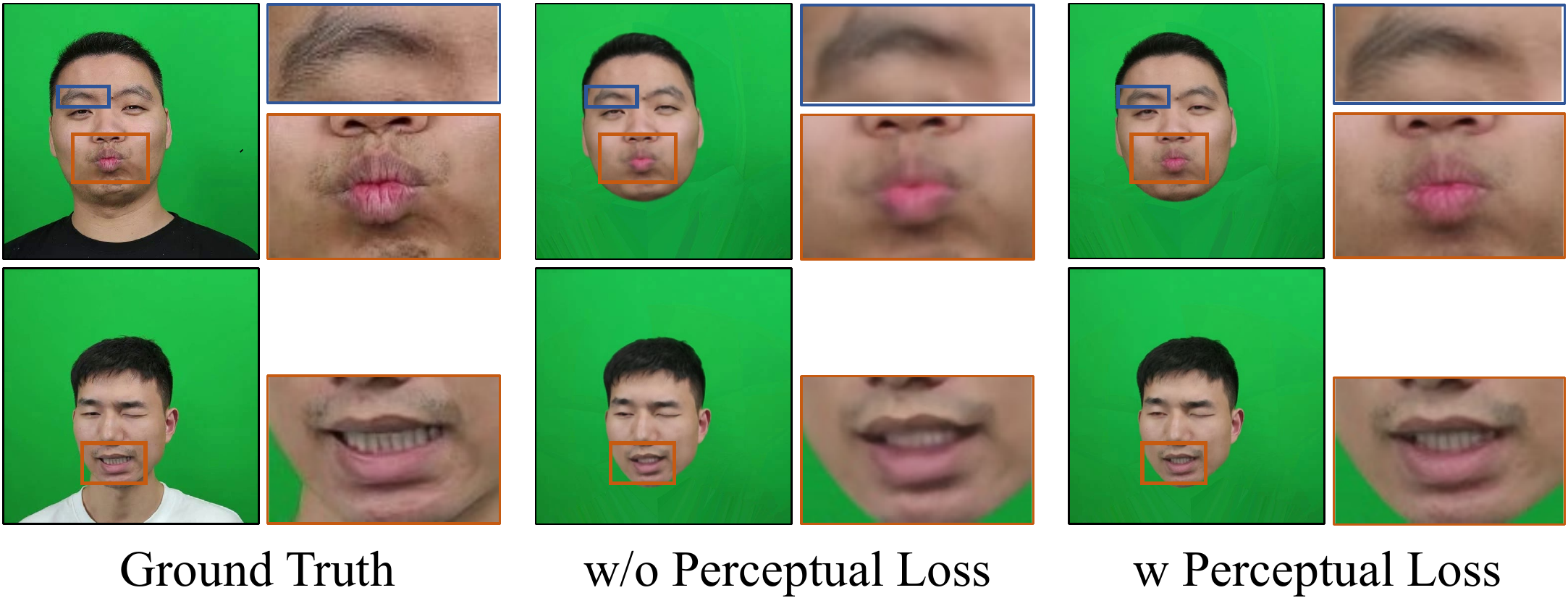}
  \caption{Ablation study on Perceptual Loss}
  \label{fig:abl_loss}
  \Description{baseline}
\end{figure}

Fig.~\ref{fig:abl_loss} shows the comparison between the results with/without perceptual loss supervision. It can be observed that the perceptual loss effectively improves the rendering quality and personalized facial attributes.
This gain comes from the fact that the perceptual loss effectively maintains the visual similarity between the predicted image and the ground truth by minimizing the distance of the two images' features extracted by a pre-trained model.

\subsection{Discussion on Expression-Aware Density Grid Update}

\begin{figure}[htb]
  \centering
  \includegraphics[width=0.95\linewidth]{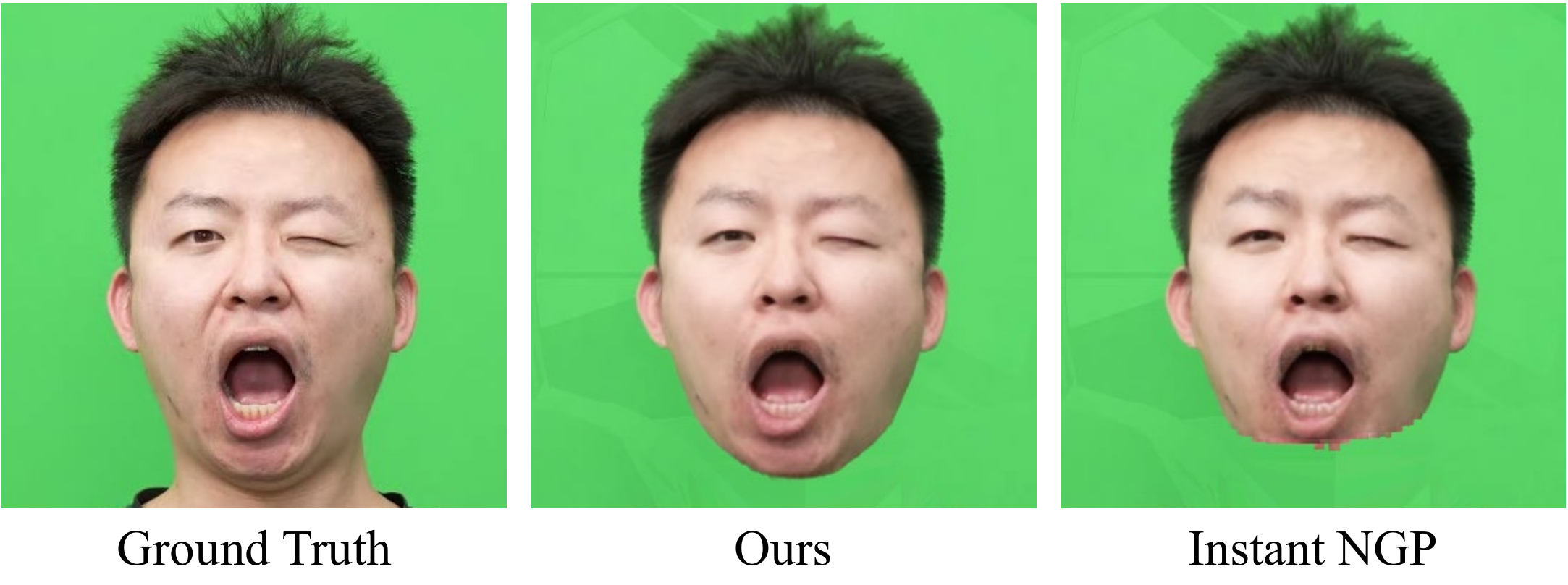}
  \caption{Ablation study on expression-aware density grid update strategy.}
  \label{fig:grid}
\end{figure}

\begin{figure}[htb]
  \centering
  \includegraphics[width=0.8\linewidth]{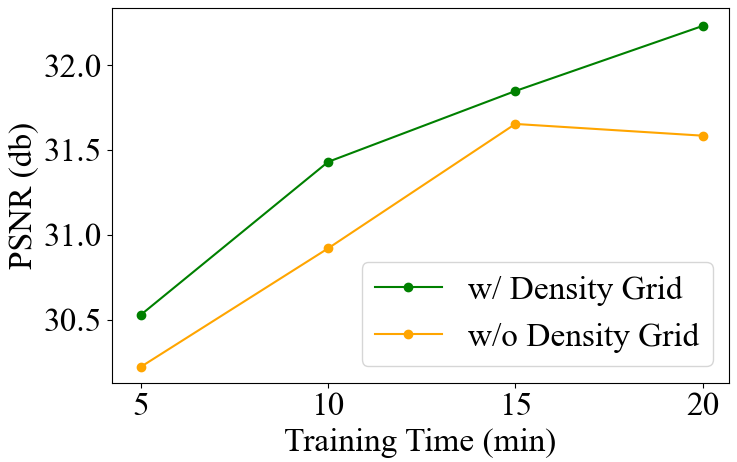}
  \caption{Ablation study on training with/without density grid. Within the same training time, the rendering quality can be better with density grid.}
  \label{fig:grid2}
\end{figure}

The density grid could mark the spatial occupation of the density field and indicate the empty area to be skipped during raymarching, which could accelerate the computation and reduce hash collisions. Therefore, an effective grid update strategy is necessary for constructing our model. Unlike the original density grid from Instant NGP~\cite{mueller2022instant}, where a static scene is considered, we need to handle a dynamic head scene. To this end, we consider more expressions of the training dataset and adopt an expression-aware way to prune the density grid. As shown in Fig.~\ref{fig:grid}, the proposed update strategy effectively helps our model generate more reasonable results. In contrast, it is difficult to deal with some expressions like mouth open if we only use a static neutral head to instruct density grid update. The training processes using/not using a density grid are shown in Fig.~\ref{fig:grid2}. We could see that a model could achieve better PSNR when a density grid
is used.

\section{Limitations}
Similar with other NeRF based face modeling approaches like NerFACE and NHA, artifacts may occur in some local regions when extrapolating the expression coefficients to a value that is far from the training distribution. This problem might be circumvented by explicitly modeling the underlying geometry like some NeRF Editing approaches\cite{neumesh,Yuan22NeRFEditing}, and we leave this as a future work.

The camera parameters and input conditions are important for NeRF based techniques. Large errors in tracking may cause losing details in our constructed model.

Thanks to the subject-specific NeRF training, our method works well for different genders, skin colors, accessories, and hairstyles. However, if the hair is in a fast and heavy non-rigid deformation, artifacts may happen in hair region due to the lack of non-rigid deformation condition.

\section{Conclusion}
We have proposed a personalized semantic facial NeRF model, which could reconstruct a subject-specific 4D avatars with only minutes of monocular RGB video. Compared with mesh-based blendshape model, our model could generate photo-realistic results and model much more personalized facial attributes, such as hair, wearings, and even muscle movements. Meanwhile, our model has better representation ability and is easy for the MLP-based implicit function to learn the dynamic head. Compared with other NeRF-based parametric head models, our model can be constructed with much less time, has better rendering quality, and contains rich facial details. It is also worth pointing out that our implicit linear blending architecture can be adopted in other problems where a linear combination relationship plays an important role due to its representation ability and training efficiency. We believe our work is a forward step toward the future digital human.

\begin{acks}
This research was supported by the National Natural Science Foundation of China (No.62122071, No.62272433), and the Fundamental Research Funds for the Central Universities (No. WK3470000021).
\end{acks}


\bibliographystyle{ACM-Reference-Format}
\bibliography{sample-base}

\end{document}